\documentclass[a4paper, 12pt, oneside]{article}

\usepackage{dsfont, amsmath, amsfonts, amssymb, mathrsfs, mathcomp, stmaryrd, graphicx}
\usepackage[top=2.5cm, left=2.5cm, right=2.5cm, bottom=2.5cm]{geometry}
\usepackage{makeidx}
\usepackage{hyperref}
\hypersetup{colorlinks=false}

\makeindex

\begin{document}

%
%
\begin{center}
\section*{Lagrangian formulation of Newtonian cosmology (Formula\c c\~{a}o lagrangiana da cosmologia newtoniana)}
%
%
H. S. Vieira$^{1}$ and V. B. Bezerra$^{1}$\\
$^{1}$Departamento de F\'{i}sica, Universidade Federal da Para\'{i}ba, Caixa Postal 5008, CEP 58051-970, Jo\~{a}o Pessoa, PB, Brazil\\
email: horacio.santana.vieira@hotmail.com and valdir@fisica.ufpb.br
\end{center}
%
%
In this paper, we use the Lagrangian formalism of classical mechanics and some assumptions to obtain cosmological differential equations analogous to Friedmann and Einstein equations, obtained from the theory of general relativity. This method can be used to a universe constituted of incoherent matter, that is, the cosmologic substratum is comprised of dust.\\
%
%
\textbf{Keywords:} total energy, cosmological equations, dust cloud.\\
%
%
Neste artigo, usamos o formalismo Lagragiano da mec\^{a}nica cl\'{a}ssica e algumas hip\'{o}teses para obter as equa\c c\~{o}es diferenciais cosmol\'{o}gicas an\'{a}logas as equa\c c\~{o}es de Friedmann e Einstein, obtidas a partir da teoria da relatividade geral. Este m\'{e}todo pode ser usado para um universo constitu\'{i}do de mat\'{e}ria incoerente, isto \'{e}, o substrato cosmol\'{o}gico \'{e} composto de poeira.\\
%
%
\textbf{Palavras-chave:} energia total, equa\c c\~{o}es cosmol\'{o}gicas, nuvem de poeira.
%
%
\section{Introduction}
The modern cosmology, relativistic or Einsteinian, is described using the theory of general relativity, whose formulation is geometric, so that the descriptions of the cosmological effects are associated to spacetime geometry. This formulation makes use of the spacetime concept which is worked out through the differential manifold concept. Others themes of modern mathematics, of complex nature, such as tensor algebra and continuous groups, are also used in this formulation \cite{d'Inverno:1998}.

Initially, the Einsteinian cosmology was not accepted to describe the universe, constituting simply in one more new cosmological theory. At the beginning, more exactly in the year of 1917, two solutions were found: one by Netherlander astronomer Willem de Sitter, and other by Albert Einstein. The first was of a universe with matter, while the second led to existence of an empty universe. Other solutions were found by Aleksandr Friedmann \cite{ZPhys.10.377} and Georges Lemaître \cite{AnnSocSciBruxA.53.51}, which gone unnoticed until the discover of the American astronomer Edwin Hubble concerning the cosmological Doppler effect by examining the light from distant stars. Hubble's discovery led cosmologists to conclude that the redshift of the emitted light by these stars could be associated to the fact that the universe would expand.

In these first years of the 1930's, the Einsteinian cosmology began to be accepted, now with less restrictions, in the sense that it predicted a model compatible with the astronomical observations. However, in 1934, Milne and McCrea \cite{QJMath.5.64,QJMath.5.73} adopted an approach based on Newtonian theory, in which it has no sense to associate the gravitational phenomena to the effects of the spacetime curvature. In this context, was shown that the universe behavior could be understood on the basis of classical physics, which does not use the mathematical complexity in the study of the universe, as is the case of Einsteinian cosmology. This mean, among other things, that is possible, in this scenario, get back the same results provided by homogeneous and isotropic models of the universe, in a simpler way from the mathematical point of view.

In Milne approach, called Newtonian cosmology, the expansion of the universe was not something dynamic, inherent to the universe itself. In this, the universe is static. However, it was necessary to incorporate the Hubble observations about the expanding universe. The solution was to admit that the observed expansion is associated to the motion of particles (galaxies) in the universe. Therefore, this particles motion in a static space would produce the same phenomena as those generated by stationary particles in the expanding universe. Thus, the expansion was understood as being caused by the movements of the particles, and not the space, allowing preserve Euclidean geometry, thus not having the need to introduce the curved spacetime of the relativistic approach. Newtonian cosmology was originally formulated for null pressure. Some decades later, the pressure term was included \cite{AnnPhys.35.437,RevModPhys.39.862}. We can introduce also in the Newtonian approach to cosmology a term containing the cosmological constant, associated with a kind of cosmological force. This term, proposed by Einstein, has been the subject of several studies, especially in recent years, with the aim to explain the question related to the accelerated expansion of the Universe.

As shown by Milne and McCrea \cite{QJMath.5.64,QJMath.5.73}, in Newtonian cosmology the cosmological equation is obtained from the equation of motion for particles (galaxies) submitted to gravitational forces \cite{BolSocAstronBras.14.34,AmJPhys.73.653}. This leads us immediately to an idea of use the formalism developed by Lagrange and Hamilton, since such it provides us the equation and integral of motion, respectively.

This paper is organized as follows. In section 2 we present the cosmological principle in the Newtonian version. In section 3 we introduce some assumptions about the structure of the universe, and then we obtain the analogous of the Einstein equation for the scale parameter. In section 4 we obtain the analogous of the Friedmann equation. Finally, in section 5 we present our conclusions.
%
%
\section{Cosmological principle}
The cosmology is based in a principle of simplicity, which is the generalization of Copernican principle. It afirms that: in each epoch, the universe presents the same aspect in all points, except by local irregularities. Thus, our universe in a certain determined Newtonian time, $t=\mbox{constant}$, is isotropic and homogeneous.
%
%
\section{Cosmological equation}
We admit as hypothesis that the cosmological substratum is made by a gas cloud in expansion, in an arbitrarily large volume, however finite. The particles that constitute this gas are the galaxies. We assume that the pressure in the cloud will be given by $p=p(t)$, which in the simplest cosmological models it is assumed to be null, which implies that the cosmological gas cloud is really a dust cloud (incoherent matter).

Two basic equations of cosmology govern the expansion of the universe. They can be understood as statements about energy, since there is a constant like the sum of kinetic and gravitational energy for the motion of a galaxy in the expansion. The expansion of the universe is observed with the relative motion of the galaxies. The universe is like an expanding gas, but the units are galaxies, and an individual galaxy does not expand.

The Hubble law describes what is observed. The speeds of galaxies moving away from us are proportional to their distances from us. Galaxies at distance $R$ are moving away from us with average speed
\begin{equation}
\frac{dR}{dt} \equiv \dot{R}=HR\ ,
\label{eq:Hubble}
\end{equation}
where $H=H(t)$ is the Hubble parameter. The Hubble law is what we expect to observe in a spatially isotropic uniformly expanding universe. We assume it would be the same for observers at any other location, and that the universe is, in fact, homogeneous, or in other words, the same everywhere and at any given time.

Gravity pulls the galaxies together and slows the expansion of the universe. If distances are measured from a typical galaxy, which could be at any location, the force of gravity on a galaxy at distance $R$ coming from the mass of homogeneous universe inside the sphere of radius $R$ is the same as if all the mass were at the center of the sphere. There is no force arising from the the region outside to the sphere.

The kinetic energy for the motion of a galaxy expressed in fixed rectangular coordinates (comoving) is a function only of $\dot{R}$ and, if the galaxy moves in a conservative force field, the potential energy is a function only of $R$. Thus, we can write
\begin{equation}
T=T(\dot{R})\ ,
\label{eq:cinetica}
\end{equation}
\begin{equation}
U=U(R)\ .
\label{eq:potencial_gravitacional}
\end{equation}

With the usual expressions for the kinetic and gravitational potential energies, the Lagrangian $\mathcal{L}=\mathcal{L}(q,\dot{q})$ for the motion of a galaxy of mass $m$ in the expansion is
\begin{equation}
\mathcal{L}(R,\dot{R})=T-U=\frac{1}{2}m\dot{R}^{2}+\frac{GMm}{R}\ ,
\label{eq:antiga_Lagrangeana}
\end{equation}
where we choose the generalized coordinates $q=R$ and $\dot{q}=\dot{R}$. In Eq.~(\ref{eq:antiga_Lagrangeana}) $M$ is the total mass of the universe, that we assume to be distributed in a sphere of radius $R$.

If we assume that there is a cosmological force on a particle (galaxy) of the gas, given by \cite{BolSocAstronBras.14.34}
\begin{equation}
F_{\Lambda}=\frac{1}{3}\Lambda mR\ ,
\label{eq:forca_cosmologica}
\end{equation}
where $\Lambda$ is the cosmological constant, we have an additional potential energy given by
\begin{equation}
U_{\Lambda}=-\int_{0}^{R}F_{\Lambda}dR'=-\frac{1}{6}\Lambda mR^{2}\ .
\label{eq:potencial_adicional}
\end{equation}
This implies that the Lagrangian, given by Eq.~(\ref{eq:antiga_Lagrangeana}), changes to
\begin{equation}
\mathcal{L}(R,\dot{R})=T-U_{eff}=T-U-U_{\Lambda}=\frac{1}{2}m\dot{R}^{2}+\frac{GMm}{R}+\frac{1}{6}\Lambda mR^{2}\ .
\label{eq:Lagrangeana}
\end{equation}
Now, using the Euler-Lagrange equation, namely
\begin{equation}
\frac{\partial \mathcal{L}}{\partial q}-\frac{d}{dt}\left(\frac{\partial \mathcal{L}}{\partial\dot{q}}\right)=0\ ,
\label{eq:Euler-Lagrange}
\end{equation}
we have:
\begin{equation}
\frac{\partial \mathcal{L}}{\partial R}=-\frac{GMm}{R^{2}}+\frac{1}{3}\Lambda m R\ ;
\label{eq:movimento_1}
\end{equation}
\begin{equation}
\frac{\partial \mathcal{L}}{\partial \dot{R}}=m\dot{R}\ \Rightarrow\ \frac{d}{dt}\left(\frac{\partial \mathcal{L}}{\partial \dot{R}}\right)=m\ddot{R}\ .
\label{eq:movimento_2}
\end{equation}
Substituting Eqs.~(\ref{eq:movimento_1}) and (\ref{eq:movimento_2}) into Eq.~(\ref{eq:Euler-Lagrange}), we obtain the equation of motion
\begin{equation}
\ddot{R}=-\frac{GM}{R^{2}}+\frac{1}{3}\Lambda R\ .
\label{eq:movimento}
\end{equation}
Since the mass of the sphere is given by
\begin{equation}
M=\frac{4}{3}\pi R^{3}\rho\ ,
\label{eq:massa_da_esfera}
\end{equation}
where $\rho$ is the mass density, Eq.~(\ref{eq:movimento}) becomes
\begin{equation}
\ddot{R}=-\frac{4}{3}\pi G\rho R+\frac{1}{3}\Lambda R\ .
\label{eq:cosmologica}
\end{equation}
Therefore, this is the Newtonian cosmological equation for the scale parameter $R$ that governs the universe expansion. This equation is analogous to Einstein equation obtained from theory of general relativity \cite{d'Inverno:1998}, in the case $p=0$, that is, dust cloud.
%
%
\section{Cosmological differential equation}
According our previous assumption the time is homogeneous within an inertial reference frame. Therefore, the Lagrangian that describs a closed system, i.e., a system not interacting with anything outside the system, cannot depend explicitly on the time. In our case, the Lagrangian is likewise independent of the time, because the system is under the action of a uniform force field. Thus, the constant quantity of the motion is $\mathcal{H}$, called Hamiltonian of the system, which can be defined as
\begin{equation}
\mathcal{H}=\left(\frac{\partial \mathcal{L}}{\partial\dot{q}}\right)\dot{q}-\mathcal{L}\ .
\label{eq:Hamiltoniano}
\end{equation}
Using the Eq.~(\ref{eq:Lagrangeana}), we obtain
\begin{equation}
E=\left(m\dot{R}\right)\dot{R}-\frac{1}{2}m\dot{R}^{2}-\frac{GMm}{R}-\frac{1}{6}\Lambda mR^{2}\ ,
\label{eq:energia}
\end{equation}
where we renamed the constant $\mathcal{H}$ by $E$, total energy, which is the constant of the motion for this case. Identifying some constants in Eq.~(\ref{eq:energia}), it can be rewritten as
\begin{equation}
\dot{R}^{2}=\frac{C}{R}+\frac{1}{3}\Lambda R^{2}-k\ ,
\label{eq:dif_cosmologica}
\end{equation}
where $C=8 \pi G \rho R^{3}/3$ and $k=-2E/m$ are constants. Therefore, this is the cosmological differential equation for the scale parameter $R$ that governs the universe expansion. This equation is analogous to Friedmann equation obtained from theory of general relativity \cite{d'Inverno:1998}.
%
%
\section{Conclusions}
In this paper, we presented a new method to obtain the cosmological differential equations, which are analogous to Friedmann and Einstein equations, in the case of null pressure. This method is equivalent to one developed by Milne and McCrea in the sense that we started from statements about the total energy for a system of particles (galaxies) which constitute the cosmological substratum.

The Newtonian approach is quite simple, from the conceptual and mathematical point of views. Thus, the approach based on classical mechanics permits to obtain in a simple way an explanation about some aspects of the universe we observe today.

The description of natural phenomena addressed by modern cosmology can, therefore, be investigated within a purely classical perspective, using the flat and static space, the Newtonian time, the dynamics of Lagrange and Hamilton, and the Newton's law of universal gravitation, plus some ad-hoc hypotheses. This approach avoids the mathematical complexity needed in the Einsteinian description, and provides the same relativistics results, within certain limitations, naturally.
%
%
\section*{Acknowledgments}
The authors would like to thank Conselho Nacional de Desenvolvimento Cient\'{i}fico e Tecnol\'{o}gico (CNPq) for partial financial support.
%
%

%
%
\end{document}